\input{aipcheck}
 
\documentclass[
    ,final            % use final for the camera ready runs
  ]
  {aipproc}

\usepackage{bm}

\newcommand{\be}{\begin{equation}}
\newcommand{\bea}{\begin{eqnarray}}
\newcommand{\ee}{\end{equation}}
\newcommand{\eea}{\end{eqnarray}}
\newcommand{\bpi}{\begin{picture}}
\newcommand{\bce}{\begin{center}}
\newcommand{\epi}{\end{picture}}
\newcommand{\ece}{\end{center}}

\def\gb{\bm{\Gamma}}

\layoutstyle{8x11double}

\begin{document}

\title{Nonperturbative gluon and ghost propagators in $d=3$}

\classification{12.38.Lg,12.38.Aw,12.38.Gc }
\keywords      {Gluon and ghost propagators, Schwinger-Dyson equations, 
dynamical gluon mass generation}

\author{Joannis Papavassiliou}{
  address={Department of Theoretical Physics and IFIC, 
University of Valencia-CSIC, E-46100, Valencia, Spain.}
}

\begin{abstract}
 
We study  the 
nonperturbative gluon and ghost propagators in $d=3$ Yang-Mills, 
using the Schwinger-Dyson  equations of the pinch technique. 
The use of the Schwinger mechanism leads to 
the dynamical generation of a gluon mass, 
which, in turn, gives rise to an infrared finite gluon propagator and ghost dressing function. 
The propagators obtained are in very good agreement with the results of $SU(2)$ lattice simulations.

\end{abstract}

\maketitle

%%%%%%%%%%%%%%%%%%%%%%%%%%%%%%%%%%%%%%%%%%%%
%% MAINMATTER
%%%%%%%%%%%%%%%%%%%%%%%%%%%%%%%%%%%%%%%%%%%%

\section{Introduction}

Even though $\rm{QCD}_3$ differs from $\rm{QCD}_4$ in several aspects,  
both theories share a crucial nonperturbative property: they 
cure their infrared (IR) instabilities through the dynamical generation of a 
gauge boson (gluon) mass, 
 without affecting    the   local gauge invariance,    which   remains
intact~\cite{Cornwall:1982zr,Alexanian:1995rp}.    
The nonperturbative dynamics  that gives rise to the generation of such 
a mass can be ultimately traced back to a subtle realization 
of the Schwinger mechanism~\cite{Schwinger:1962tn,Jackiw:1973tr}. 
The  gluon  mass generation  manifests  itself  at  the level  of  the
fundamental Green's  functions of the  theory in a very  distinct way,
giving rise to an IR behavior that would be difficult to explain
otherwise.  Specifically,  in the Landau gauge, both  in $d=3,4$, 
the gluon  propagator and the  ghost dressing function reach  a finite
value  in the deep  IR~\cite{Aguilar:2008xm, Boucaud:2008ji,Boucaud:2010gr,Aguilar:2010zx}. 
However, the gluon
propagator of $\rm{QCD}_3$ displays  a local maximum at relatively low
momenta~\cite{Cucchieri:2003di,Quandt:2010yq}, before reaching a  finite value at $q=0$. 
This characteristic
behavior is  qualitatively different to what happens  in $d=4$, where
the gluon  propagator is a monotonic  function of the  momentum in the
entire range between the IR and UV fixed points~\cite{Aguilar:2008xm}.

Given that the gluon mass generation 
is a purely nonperturbative effect,
it can be naturally treated within the framework  
of the Schwinger-Dyson equations (SDE). 
 These complicated dynamical equations are best studied in
a  gauge-invariant framework based on the  pinch technique  
(PT)~\cite{Cornwall:1982zr,Cornwall:1989gv,Binosi:2009qm}, and its profound  
correspondence with the background field method (BFM)~\cite{Abbott:1980hw}.
As has been explained in detail in the recent literature~\cite{Aguilar:2006gr}, 
this latter formalism  
allows for a  gauge-invariant truncation of the SD series, 
in the sense that it preserves manifestly, and at every step,  
the transversality of the gluon self-energy.

In the present talk we report on a recent study of the gluon and ghost propagators 
of pure Yang-Mills in $d=3$, using the SDEs  
of the PT-BFM formalism in the Landau gauge~\cite{Aguilar:2010zx} 
(for different approaches see, e.g.,\cite{Dudal:2008rm,Braun:2007bx}).

\section{Gluon Mass generation in Yang-Mills theories}

In order to understand the basic concept underlying the Schwinger mechanism, 
let us consider the gluon propagator (in the Landau gauge),   
\be 
\Delta_{\mu\nu}(q)= -i {\rm P}_{\mu\nu}(q)\Delta(q^2)\,,
\label{fprop}
\ee
where 
${\rm P}_{\mu\nu}(q)= g_{\mu\nu} - q_{\mu} q_{\nu}/q^2$.
The scalar factor $\Delta(q^2)$ is given by 
$\Delta^{-1}(q^2) = q^2 + i \Pi(q^2)$,  
where  \mbox{$\Pi_{\mu\nu}(q)={\rm P}_{\mu\nu}(q) \,\Pi(q^2)$} 
is the gluon self-energy.  
One usually defines 
the dimensionless vacuum polarization, to be denoted by ${\bm \Pi}(q^2)$, 
as 
$\Pi(q^2)=q^2 {\bm \Pi}(q^2)$, and thus  
\be
\Delta^{-1}(q^2) = q^2 [1 + i{\bm \Pi}(q^2)]\,.
\label{funcbar}
\ee 
As Schwinger pointed out long time ago~\cite{Schwinger:1962tn},     
the gauge invariance of a vector field does not necessarily 
imply zero mass for the associated particle, if the 
current vector coupling is sufficiently strong. 
According to Schwinger's fundamental observation, 
if ${\bm \Pi}(q^2)$ 
acquires a pole at zero momentum transfer, then the 
 vector meson becomes massive, even if the gauge symmetry 
forbids a mass at the level of the fundamental Lagrangian.
Indeed, it is clear that if the vacuum polarization ${\bm \Pi}(q^2)$ 
has a pole at  $q^2=0$ with positive residue $m^2$, i.e.,  
\be
{\bm \Pi}(q^2) = m^2/q^2,
\ee
then (in Euclidean space)
\be
\Delta^{-1}(q^2) = q^2 + m^2.
\ee
Thus, the vector meson 
becomes massive, $\Delta^{-1}(0) = m^2$, 
even though it is massless in the absence of interactions ($g=0$). 
There is {\it no} physical principle that would preclude ${\bm \Pi}(q^2)$ from 
acquiring such a pole, even in the absence of elementary scalar fields. 
In a {\it strongly-coupled} theory, like nonperturbative Yang-Mills in \mbox{$d=3,4$}, 
this may happen for purely dynamical reasons,
since strong binding may generate zero-mass 
bound-state excitations~\cite{Jackiw:1973tr}.
The latter  act  {\it  like}
dynamical Nambu-Goldstone bosons, in the sense that they are massless,
composite,  and {\it longitudinally   coupled};  but, at  the same  time, they
differ  from  Nambu-Goldstone  bosons   as  far  as  their  origin  is
concerned: they  do {\it not} originate from  the spontaneous breaking
of  any global symmetry~\cite{Cornwall:1982zr}.
 In what follows we will assume that the theory can  
indeed generate the required bound-state poles;  
the demonstration of the existence of such bound states  
is a difficult dynamical problem, that must be addressed 
by means of Bethe-Salpeter equations.

%%%%%%%%%%%%%%%%%%%%%%%%%%%%%%%%%%%%%%%%%%%%%%%%%%%
%          Figure 
%%%%%%%%%%%%%%%%%%%%%%%%%%%%%%%%%%%%%%%%%%%%%%%%%%%%
\begin{figure}[!b]
\resizebox{0.9\columnwidth}{!}
{\includegraphics{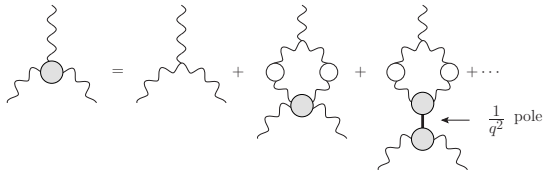}}
\caption{Vertex with  nonperturbative massless excitations triggering the Schwinger mechanism.}
\label{bound_new}
\end{figure}
%%%%%%%%%%%%%%%%%%%%%%%%%%%%%%%%%%%%%%%%%%%%%%%%%%%

The Schwinger mechanism
is incorporated into the SDE of the gluon propagator 
through the form of the nonperturbative three-gluon vertex (Fig.\ref{bound_new}). 
In fact, in order for the gauge symmetry to be preserved, the three-gluon 
vertex must satisfy the same Ward identity as in the massless case, 
but now with massive, as opposed to 
massless, gluon propagators on its rhs.   
The way this crucial requirement is enforced is precisely through the incorporation into the 
three-gluon vertex of 
the Nambu-Goldstone (composite) massless excitations mentioned above. 
To see how this works with a simple example, 
let us consider the standard tree-level vertex 
\be
\Gamma_{\mu\alpha\beta}(q,p,r) = (q-p)_{\beta}g_{\mu\alpha} 
+ (p-r)_{\mu}g_{\alpha\beta} + (r-q)_{\alpha}g_{\mu\beta}\,,
\label{stanvert}
\ee
which satisfies the simple Ward identity 
\be
q^{\mu}\Gamma_{\mu\alpha\beta}(q,p,r) = 
P_{\alpha\beta}(r) \Delta_0^{-1}(r) - P_{\alpha\beta}(p) \Delta_0^{-1}(p)
\label{ewi}
\ee
where $\Delta_0^{-1}(q) = q^2$ is the inverse of the tree-level propagator. 
After the dynamical mass generation, the inverse gluon propagator becomes, roughly speaking,  
\be
\Delta_m^{-1}(q^2)= q^2 - m^2(q^2), 
\label{rg5}
\ee
and the new vertex, $\gb_{\mu\alpha\beta}^{m}(q,p,r)$ that must replace 
$\Gamma_{\mu\alpha\beta}(q,p,r)$ must still satisfy the Ward identity of (\ref{ewi}), but with 
 $\Delta_0^{-1} \to \Delta_m^{-1}$ on the rhs.
This is accomplished if   
\be
\gb^{m}_{\mu\alpha\beta}(q,p,r) = \Gamma_{\mu\alpha\beta}(q,p,r) + 
V_{\mu\alpha\beta}(q,p,r),
\label{verv}
\ee
where $V_{\mu\alpha\beta}(q,p,r)$ contains the massless poles. A standard Ansatz for 
 $V_{\mu\alpha\beta}(q,p,r)$ is~\cite{Cornwall:1984eu} 
\bea
V_{\mu\alpha\beta}(q,p,r) &=&
 m^2(r)\frac{q_{\mu} p_{\alpha}(q-p)_{\rho}}{2q^{2}p^{2}}\,
P^{\rho}_{\beta}(r)  
\nonumber\\
&-& 
\left[m^2(p)-m^2(q)\right]\frac{r_{\beta}}{r^{2}}\,P^{\mu}_{\rho}(q)\,P^{\rho}_{\alpha}(p)  
\nonumber\\
&+& {\rm c.p.}\,, 
\label{Ver1}
\eea
It is easy to check that 
\be
q^{\mu}V_{\mu\alpha\beta}(q,p,r) = P_{\alpha\beta}(p) m^2(p) -  P_{\alpha\beta}(r) m^2(r)\,,
\label{wiv}
\ee
and cyclic permutations.  
Therefore, one has
\be
q^{\mu}\Gamma^{m}_{\mu\alpha\beta}(q,p,r) = 
P_{\alpha\beta}(r) \Delta_m^{-1}(r) - P_{\alpha\beta}(p) \Delta_m^{-1}(p)\,,
\label{ewim}
\ee
as announced. 

%%%%%%%%%%%%%%%%%%%%%%%%%%%%%%%%%%%%%%%%%%%%%%%%%%%
%          Figure  one column
%%%%%%%%%%%%%%%%%%%%%%%%%%%%%%%%%%%%%%%%%%%%%%%%%%%%
\begin{figure}[!t]
\resizebox{0.95\columnwidth}{!}
{\includegraphics{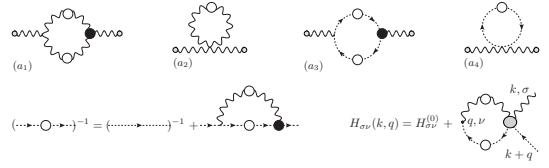}}
\caption{The SDEs for the various quantities involved.}
\label{SDeqs}
\end{figure}
%%%%%%%%%%%%%%%%%%%%%%%%%%%%%%%%%%%%%%%%%%%%%%%%%%%

%%%%%%%%%%%%%%%%%%%%%%%%%%%%%%%%%%%%%%%%%%%%%%%%%%%
%          Figure - two columns
%%%%%%%%%%%%%%%%%%%%%%%%%%%%%%%%%%%%%%%%%%%%%%%%%%%%
\begin{figure}[!t]
\resizebox{.9\textwidth}{!}
{\includegraphics{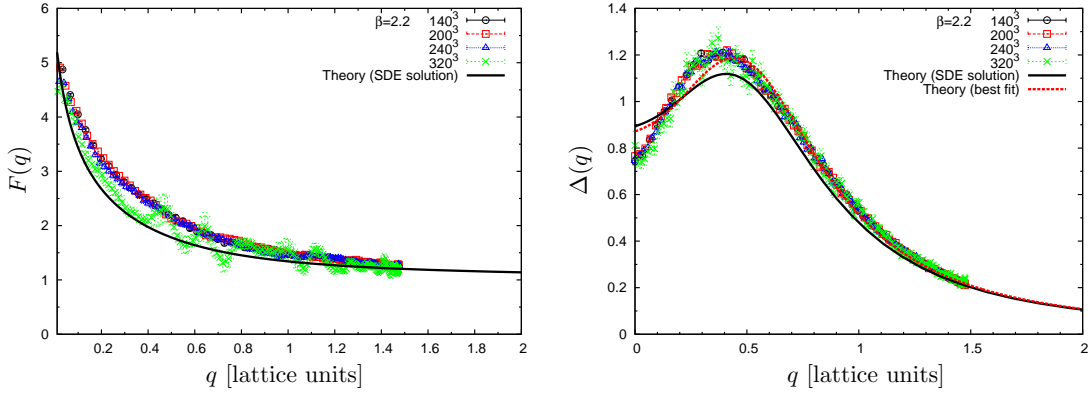}}
\caption{Comparison with the lattice results of~\cite{Cucchieri:2003di}.}
\label{prop}
\end{figure}
%%%%%%%%%%%%%%%%%%%%%%%%%%%%%%%%%%%%%%%%%%%%%%%%%%%%%

\section{SDE analysis and comparison with the lattice}

In the ``one-loop dressed'' approximation, the PT-BFM gluon self-energy is given by  
the subset of diagrams shown in Fig.\ref{SDeqs}. As explained in detail in various works (see, e.g., 
\cite{Aguilar:2006gr,Binosi:2009qm}, 
the resulting gluon self-energy is manifestly transverse, 
due to the simple Ward identities satisfied by the PT-BFM fully dressed vertices. 
In addition to the SDE of the gluon, we consider the corresponding SDEs for
(i) the ghost propagator, denoted by $D(p)$, 
(or its dressing function, $F(p)$ , given by $D(p)= iF(p)/p^2$), 
and (ii) the auxiliary function $G(q)$, defined as the $g_{\mu\nu}$
component of the function $H_{\mu\nu}$, shown in Fig.\ref{SDeqs}. 
$G(q)$
enters into the important relation 
\be
\Delta(q) = 
\left[1+G(q)\right]^2 \widehat{\Delta}(q), 
\label{bqi2}
\ee
relating the PT-BFM gluon propagator, $\widehat{\Delta}(q)$,
and the conventional one, $\Delta(q)$ (simulated on the lattice).  
The closed expressions for all these SDEs have been given in~\cite{Aguilar:2010zx}.

The way we proceed is the following. 
Instead of actually solving the system of coupled integral equation, 
we follow an approximate 
procedure, which is operationally less complicated, 
and appears to capture rather well the underlying dynamics. 

Specifically, we will assume that the PT-BFM gluon propagator has the form 
\be
\widehat\Delta^{-1}(q) = q^2 + m^2 + \widehat\Pi_{m}(q).
\label{dmp}
\ee  
where
\be
(d-1)\widehat\Pi_m(q) = [(a_1)+(a_2)+(a_3)+(a_4)]_{\mu}^{\mu}\,.
\label{ptr}
\ee
The function $\widehat\Pi_m(q)$ will be determined by calculating 
the Feynman graphs given in Fig.\ref{SDeqs},   
using inside the corresponding integrals $\Delta\to (q^2 + m^2)^{-1} $, and $D\to 1/q^2$ 
In order 
to maintain gauge invariance intact, ensuring that $q_{\mu} \widehat\Pi_{m}^{\mu\nu}(q) =0$, 
 we will use the $\gb^{m}_{\mu\alpha\beta}(q,p,r)$ 
given in  (\ref{verv}) as the fully-dressed three-gluon vertex; in 
$V_{\mu\alpha\beta}^{m}(q,p,r)$ we will use a constant (instead of a running) mass, $m$.  
As for the ghost dressing function,  
we set up an approximate version of the ghost SDE, 
and
{\it solve} it self-consistently for the unknown function $F(p)$.  
As explained in~\cite{Aguilar:2010zx}, this procedure allows for a very good 
{\it simultaneous} fit of the available lattice data. 
The best possible fit we have found is shown in Fig.\ref{prop},  
furnishing the ratio \mbox{$m/2g^2 = 0.15$}.  

\section{Conclusions}

We have presented a nonperturbative study of the 
(Landau gauge) gluon and ghost propagator for $d=3$ Yang-Mills, 
using the ``one-loop dressed'' SDEs of the PT-BFM formalism. 
One of the most powerful features of this framework is that the 
transversality of the truncated gluon self-energy is guaranteed, 
by virtue of the QED-like Ward identities satisfied by the fully-dressed vertices 
entering into the dynamical equations. 
The central dynamical ingredient of our analysis is
the assumption that the Schwinger mechanism
is indeed realized in $d=3$ Yang-Mills.
The propagators obtained from these nonperturbative 
equations agree rather well with the results of $SU(2)$ lattice simulations. 

\paragraph{Acknowledgments:
I thank the QCHS-IX organizers for their kind hospitality. This research is supported 
by the European FEDER and  Spanish MICINN under grant FPA2008-02878, and the Fundaci\'on General of the UV.}


\begin{thebibliography}{9}

\bibliographystyle{aipproc}

%\cite{Cornwall:1982zr}
\bibitem{Cornwall:1982zr}
J.~M.~Cornwall,
%``Dynamical Mass Generation In Continuum QCD,''
Phys.\ Rev.\ D {\bf 26}, 1453 (1982). 
%%CITATION = PHRVA,D26,1453;%%



%\cite{Alexanian:1995rp}
\bibitem{Alexanian:1995rp}
  G.~Alexanian and V.~P.~Nair,
  %``A Selfconsistent Inclusion Of Magnetic Screening For The Quark - Gluon
  %Plasma,''
  Phys.\ Lett.\  B {\bf 352}, 435 (1995);
 % [arXiv:hep-ph/9504256].
  %%CITATION = PHLTA,B352,435;%%
%\cite{Karabali:1998yq}
%\bibitem{Karabali:1998yq}
  D.~Karabali, C.~j.~Kim and V.~P.~Nair,
  %``On the vacuum wave function and string tension of Yang-Mills theories  in
  %(2+1) dimensions,''
  Phys.\ Lett.\  B {\bf 434}, 103 (1998);
 % [arXiv:hep-th/9804132].
  %%CITATION = PHLTA,B434,103;%%
%\cite{Buchmuller:1994qy}
%\bibitem{Buchmuller:1994qy}
  W.~Buchmuller and O.~Philipsen,
  %``Phase structure and phase transition of the SU(2) Higgs model in
  %three-dimensions,''
  Nucl.\ Phys.\  B {\bf 443}, 47 (1995).
%  [arXiv:hep-ph/9411334].
  %%CITATION = NUPHA,B443,47;%%


%\cite{Schwinger:1962tn}
\bibitem{Schwinger:1962tn}
  J.~S.~Schwinger,
  %``GAUGE INVARIANCE AND MASS,''
  Phys.\ Rev.\  {\bf 125}, 397 (1962);
  %%CITATION = PHRVA,125,397;%%
%\cite{Schwinger:1962tp}
%\bibitem{Schwinger:1962tp}
%  J.~S.~Schwinger,
  %``Gauge Invariance And Mass. 2,''
  Phys.\ Rev.\  {\bf 128}, 2425 (1962).
  %%CITATION = PHRVA,128,2425;%%

%\cite{Jackiw:1973tr}
\bibitem{Jackiw:1973tr}
  R.~Jackiw and K.~Johnson,
  %``Dynamical Model Of Spontaneously Broken Gauge Symmetries,''
  Phys.\ Rev.\ D {\bf 8}, 2386 (1973);
  %%CITATION = PHRVA,D8,2386;%%
%\cite{Cornwall:1973ts}
%\bibitem{Cornwall:1973ts}
  J.~M.~Cornwall and R.~E.~Norton,
  %``Spontaneous Symmetry Breaking Without Scalar Mesons,''
  Phys.\ Rev.\ D {\bf 8} 3338 (1973);
  %%CITATION = PHRVA,D8,3338;%%
%\cite{Eichten:1974et}
%\bibitem{Eichten:1974et}
E.~Eichten and F.~Feinberg,
%``Dynamical Symmetry Breaking Of Nonabelian Gauge Symmetries,''
Phys.\ Rev.\ D {\bf 10}, 3254 (1974).
  %%CITATION = PHRVA,D10,3254;%%

%\cite{Aguilar:2008xm}
\bibitem{Aguilar:2008xm}
  A.~C.~Aguilar, D.~Binosi and J.~Papavassiliou,
  %``Gluon and ghost propagators in the Landau gauge: Deriving lattice results
  %from Schwinger-Dyson equations,''
  Phys.\ Rev.\  D {\bf 78}, 025010 (2008).
%  [arXiv:0802.1870 [hep-ph]].
  %%CITATION = PHRVA,D78,025010;%%

%\cite{Boucaud:2008ji}
\bibitem{Boucaud:2008ji}
Ph.~Boucaud {\it et al.},
%``IR finiteness of the ghost dressing function from numerical resolution of
%the ghost SD equation,''
JHEP {\bf 0806} (2008) 012.
%[arXiv:0801.2721 [hep-ph]];

%\cite{Boucaud:2010gr}
\bibitem{Boucaud:2010gr}
  Ph.~Boucaud {\it et al.},
  %``The low-momentum ghost dressing function and the gluon mass,''
  Phys.\ Rev.\  D {\bf 82}, 054007 (2010);
%  [arXiv:1004.4135 [hep-ph]].
  %%CITATION = PHRVA,D82,054007;%%
%\cite{RodriguezQuintero:2010wy}
%\bibitem{RodriguezQuintero:2010wy}
  J.~Rodriguez-Quintero,
  %``On the massive gluon propagator, the PT-BFM scheme and the low-momentum
  %behaviour of decoupling and scaling DSE solutions,''
  arXiv:1005.4598 [hep-ph].
  %%CITATION = ARXIV:1005.4598;%%


%\cite{Aguilar:2010zx}
\bibitem{Aguilar:2010zx}
  A.~C.~Aguilar, D.~Binosi and J.~Papavassiliou,
  %``Nonperturbative gluon and ghost propagators for d=3 Yang-Mills,''
  Phys.\ Rev.\  D {\bf 81}, 125025 (2010)
%%  [arXiv:1004.2011 [hep-ph]].
%%CITATION = PHRVA,D81,125025;%%


%\cite{Cucchieri:2003di}
\bibitem{Cucchieri:2003di}
  A.~Cucchieri, T.~Mendes and A.~R.~Taurines,
  %``SU(2) Landau gluon propagator on a 140^3 lattice,''
  Phys.\ Rev.\  D {\bf 67}, 091502 (2003);
  %[arXiv:hep-lat/0302022].
  %%CITATION = PHRVA,D67,091502;%%
%\cite{Cucchieri:2010xr}
%\bibitem{Cucchieri:2010xr}
  A.~Cucchieri and T.~Mendes,
  %``Numerical test of the Gribov-Zwanziger scenario in P,''
 PoS(QCD-TNT09)026 (2010).
 % arXiv:1001.2584 [hep-lat].
  %%CITATION = ARXIV:1001.2584;%%


%\cite{Quandt:2010yq}
\bibitem{Quandt:2010yq}
  M.~Quandt, H.~Reinhardt and G.~Burgio,
  %``The role of center vortices in Gribov's confinement scenario,''
  Phys.\ Rev.\  D {\bf 81}, 065016 (2010);
%  [arXiv:1001.3699 [hep-lat]].
  %%CITATION = PHRVA,D81,065016;%%
%\cite{Burgio:2009xp}
%\bibitem{Burgio:2009xp}
  G.~Burgio, M.~Quandt and H.~Reinhardt,
  %``BRST symmetry vs. Horizon condition in Yang-Mills theories,''
  Phys.\ Rev.\  D {\bf 81}, 074502 (2010)
%  [arXiv:0911.5101 [hep-lat]].
  %%CITATION = PHRVA,D81,074502;%%



%\cite{Cornwall:1989gv}
\bibitem{Cornwall:1989gv}
  J.~M.~Cornwall and J.~Papavassiliou,
  %``Gauge Invariant Three Gluon Vertex in QCD,''
  Phys.\ Rev.\  D {\bf 40}, 3474 (1989).
  %%CITATION = PHRVA,D40,3474;%%


%\cite{Binosi:2009qm}
\bibitem{Binosi:2009qm}
  D.~Binosi and J.~Papavassiliou,
  %``Pinch Technique: Theory and Applications,''
  Phys.\ Rept.\  {\bf 479}, 1 (2009).
  %[arXiv:0909.2536 [hep-ph]].
  %%CITATION = PRPLC,479,1;%%

%\cite{Abbott:1980hw}
\bibitem{Abbott:1980hw}
L.~F.~Abbott,
%``The Background Field Method Beyond One Loop,''
Nucl.\ Phys.\ B {\bf 185}, 189 (1981).
%%CITATION = NUPHA,B185,189;%%


%\cite{Aguilar:2006gr}
\bibitem{Aguilar:2006gr}
  A.~C.~Aguilar and J.~Papavassiliou,
  %``Gluon mass generation in the PT-BFM scheme,''
  JHEP {\bf 0612}, 012 (2006)
%  [arXiv:hep-ph/0610040].
  %%CITATION = JHEPA,0612,012;%%
%

%\cite{Dudal:2008rm}
\bibitem{Dudal:2008rm}
  D.~Dudal {\it et al.},
  %``The Landau gauge gluon and ghost propagator in the refined Gribov-Zwanziger
  %framework in 3 dimensions,''
  Phys.\ Rev.\  D {\bf 78}, 125012 (2008). 
%  [arXiv:0808.0893 [hep-th]].
  %%CITATION = PHRVA,D78,125012;%%


%\cite{Braun:2007bx}
\bibitem{Braun:2007bx}
  J.~Braun, H.~Gies and J.~M.~Pawlowski,
  %``Quark Confinement from Color Confinement,''
  Phys.\ Lett.\  B {\bf 684}, 262 (2010). 
%  [arXiv:0708.2413 [hep-th]].
  %%CITATION = PHLTA,B684,262;%%


%\cite{Cornwall:1984eu}
\bibitem{Cornwall:1984eu}
  J.~M.~Cornwall, W.~S.~Hou and J.~E.~King,
  %``Gauge Invariant Calculations In Finite Temperature QCD: Landau Ghost And
  %Magnetic Mass,''
  Phys.\ Lett.\  B {\bf 153}, 173 (1985).
  %%CITATION = PHLTA,B153,173;%%


\end{thebibliography}
\end{document}